\documentclass[11pt]{article}
\usepackage{moriond,epsfig,amsmath,amssymb}

\def\be{\begin{equation}}
\def\ee{\end{equation}}
\def\bea{\begin{eqnarray}}
\def\eea{\end{eqnarray}}

\begin{document}

\begin{flushright}
\begin{tabular}{l}
IPPP/08/32\\
DCPT/08/64
\end{tabular}
\end{flushright}

\vspace*{0.0cm}
\title{Summary of hints for new physics
from (quark) flavour physics}

\author{Roman Zwicky}

\address{IPPP, Department of Physics, 
Durham University, Durham DH1 3LE, UK}

\maketitle

\vspace{0.1cm}
\begin{center}
\emph{Invited talk 44th
Rencontres de Moriond LaThuile, Italy, March 2008, \\ QCD and High Energy Interactions }
\end{center}

\vspace{0.7cm}

\abstracts{
Recently two hints for new physics have emerged:
The $B_s$ mixing phase $\phi_s$ and the rate of $D_s \to (\mu,\tau) \nu$ exposing a discrepancy 
of $\sim 3\sigma$ and $3.8\sigma $ deviation 
from the Standard Model respectively. 
Moreover the difference of the
CP asymmetries in $B \to K \pi$ between the charged and neutral modes is at the $5.3\sigma$ level which is somewhat larger than expected. 
New physics in $\phi_s$ or 
$A_{\rm CP}(B \to K \pi)$ would be in contradiction with the minimal 
flavour violation hypothesis. 
The latter has recently attracted attention 
because of the absence of
deviation in CP and flavour violation in the quark sector.}

\section*{Status of quark flavour physics \footnote{A summary on new  results in the field  has been given by Barsuk \cite{barsuk} at Moriond QCD 2008.}}

Quark flavour physics in the Standard Model (SM) is 
governed by the four parameter CKM matrix describing the 
strength of flavour transitions.  The two least known
parameters  are embedded in the normalized CKM triangle,
which has been (over)constrained
by experimental facilities like CLEO, BaBar, Belle, Tevatron, NA48 and others over the last few years.
It is well known that 
CP-violation is one of the three necessary conditions 
for Baryogenesis. In the SM the CKM sector  is the only established source of CP-violation, which is found to be insufficient (roughly ten orders of magnitude) 
in the three family case  due to small quark mass differences 
in the up and down type structures. This  renders CP-violation a 
promising territory to search for new physics (NP).
On the other hand  all the measurements from 
flavour facilities result in constraints which are
consistent with the SM, depicted 
in the by now famous plots of the CKM-fitter and
UT-fit collaborations \cite{fitter}. In short,
the CKM mechanism is self-consistent with
the current data and describes
CP-violation of $B_d$ and $K$ mesons quantitatively \footnote{CP-violation in the
$D$ and $B_s$ systems are not  established
at the $5\sigma$ level.
First CP-asymmetry measurements in the
Baryon sector have been reported at Moriond
$A_{\rm CP}(\Lambda_b^0 \to p \pi^-[K^-] ) =
0.03(17)(5)[0.37(17)(3)]$ \cite{thomas}}.
 
It might be seen as a consequence of this consistent
picture that the community has focused on rare decays and minimal flavour violation (MFV), which 
we intend to discuss briefly.

The term ``rare'' is understood to be relative to the
SM and it could be loop suppressed (flavour changing neutral currents (FCNC)), helicity suppressed (small quark mass), SU(2)$_{\rm L}$ (right-handed currents) or simply CKM suppressed (often called Cabibbo suppressed). 
The prototype of a flavour changing neutral current is the  $b \to s \gamma$, where experiment and theory are in agreement.
The potential of many observables (FCNC) 
has not been experimentally 
exhausted or not even seen yet by the 
$B$-factories BaBar and Belle and this fact is one of the many
reasons why the upgrade at Belle and a Super $B$-factory is necessary in order
to pursue the search for new physics in particle physics.
In the very near future the LHCb experiment 
\cite{dijkstra} will open the window
to the $B_s$-sector \footnote{The exciting results from the
TeVatron represent only the very beginning of $B_s$-physics.}
and reach particularly far in channels with 
muons in the final states, e.g. 
measurement of the $B_s$ mixing phase to be discussed below,
$B_{s} \to \mu\mu$, $B_s \to \phi \gamma$\;\cite{MXZ} and $B_{d(s)} \to K^*(\phi) \mu\mu$  \cite{susy}.


The idea behind the minimal flavour hypothesis is that the
only sources of violation of the global flavour symmetry $G_F = U(3)^5$ are the Yukawa matrices. This symmetry can be formally recovered 
by promoting the Yukawas to spurious fields transforming
accordingly under $G_F$. This allows  MFV to be defined 
as an effective field theory \cite{gino}, which is formally 
invariant under $G_F$.  
The theory is minimal in the sense that the Yukawas 
remain the only source of flavour violation.
In the original paper no new sources of CP-violation 
were assumed, but within the MSSM for instance
certain authors have allowed for new CP phases 
which show up in dipole operators. 
The concept of MFV is appealing, experimentally 
motivated and testable because it predicts correlations among observables due to a small number of possible operators. 
An interesting result is for instance that MFV imposed on 
the R-parity violating MSSM evades the bounds from 
Proton decay \cite{christopher} and in this connection is 
a viable alternative to R-parity itself.

A model generating
a MFV structure is basically equivalent 
to a model explaining the masses and flavour 
violations. This  has proven to be a  notoriously difficult area where simple ideas are  sparse.
Furthermore if the spurious Yukawa matrices
are thought to be degrees of freedom acquiring
a vacuum expectation value then some of them
correspond to Goldstone bosons and require
a yet unknown mechanism for generating their mass.
In order to assess  the large hierarchy between
the top Yukawa as compared to the other flavours
it was recently proposed \cite{ThomasThorsten}
to consider the consecutive breaking of the quark
flavour group 
$G_F|_\Lambda \stackrel{}{\to} SU(2)_{Q_L} \times
 SU(2)_{U_R}  \times SU(3)_{D_R} \times .. 
 |_{\Lambda'}\stackrel{}{\to} 1$   \footnote{We are 
taking a cavalier attitude to additional $U(1)$
factors above.} at the  scales $\Lambda$
and $\Lambda'$ 
such that $\Lambda/\Lambda' \sim m_t/m_b$.

\subsection*{Non-standard $D_s \to l \nu$ decays? \cite{fds}}

In the SM the rate of leptonic decays  as $D_s \to l \nu$ 
is simply given by
\begin{equation}
\Gamma(D_s \to l \nu) \sim 
|V_{\rm cs}|^2 f_{D_s}^2 \cdot G_F^2   m_l^2 m_{D_s} \Big(1-\frac{ m_l^2}{m_{D_s}^2}\Big)^2 \,,
\end{equation}
exposing the famous helicity suppression factor 
$(m_l/m_{D_s})^2$ familiar from $\pi \to  l \nu$.
This helicity suppression is relieved in decays
$D_s \to D_s^* (\gamma) \to l \nu (\gamma)$ but should  not 
contaminate the process $l  = \mu$ by more than one percent
and is entirely negligible for $l = \tau$ 
\cite{Daniel,Sheldon,Andreas}.
In practice the product $f_{D_s} |V_{\rm cs}|$ is constrained and if one
quantity is known it allows us to determine the other one.

Initially the plan was to validate
the precision of lattice QCD in the semileptonic charm sector in order to gain confidence for predictions in the $B$-sector and to extract 
$|V_{\rm ub}|$ from $B \to \tau \nu$ with predictions for $f_B$ for instance. 

The story is now a different one. 
The HPQCD collaboration   has 
provided precise predictions,
\begin{equation}
\label{eq:lat}
f^{HPQCD}_{D_s} =  241(3) \, {\rm MeV}  \,,     \qquad         f^{HPQCD}_{D^+} =  208(4) \, {\rm MeV}  \,,
\end{equation}
with 2+1 staggered fermions (and therefore unquenched) \cite{hpqcd}.
Averaged measurements from CLEO, BaBar and Belle \cite{fdsexp}
with the SM relation $|V_{\rm cs}| \simeq |V_{\rm ud}| = 0.97377(27)$ leads to
a decay constant
\begin{equation}
\label{eq:lattice}
f_{D_s}^{exp} \; = \;  \left\{ \begin{array}{l}  
274(10) \, {\rm MeV}  \mbox{ \cite{Sheldon}}  \\
277(09) \, {\rm MeV}  \mbox{ \cite{Andreas}} \\
\end{array}  \right. \quad ,
\end{equation}
where the first average excludes certain measurements with
potentially problematic normalizations to the $D_s \to \pi^+ \phi$ mode.
The experimental and HPQCD predictions differ
by more than $3 \sigma$, in fact $3.8\sigma$ in the analysis presented in \cite{Andreas}. 
The 2 Higgs Doublet Model (2HDM) of type II, most prominently embedded in the MSSM, is an example of a model which is sensitive to this channel.
The decay rate gets modified by a multiplicative 
factor $[1-(\tan \beta \,  m_{D_s}/m_{H_+})^2 (m_s/(m_c+m_s)]^2$, which shows destructive interference and therefore
has the wrong characteristic to explain the effect\,\footnote{C.f. reference\,\cite{Andrew} for a first numerical analysis in conjunction with the MSSM and the CLEO-c experiment.}.
There 
is also a term from the right-handed coupling
of the up-type quarks proportional to
$m_U$,  which adds
constructively but is negligible due   
to a suppression factor $(m_{D_s}/m_{H_+})^2$ (independent of $\tan\beta$ in the rate), at least
in more standard 2HDM's\,\cite{Andreas}.
In general though the helicity suppression could be relieved in other models of NP, but the separate average of the 
$\mu$ and $\tau$ channels clearly contradicts  
\begin{equation}
 f^{exp\,(\mu\nu)}_{D_s} =  273(11) \, {\rm MeV}  \,,     \qquad         f^{exp\,(\tau\nu)}_{D_s} =  285(15) \, {\rm MeV}  \mbox{ \cite{Andreas}} \,, 
\end{equation}
this hypothesis as it is the $\tau$ and not the $\mu$ channel which
leads to a larger decay constant! The situation has become even
more interesting as very recently CLEO has released data 
on  $D_+ \to \mu_+ \nu$ \cite{CLEOnew} from where they extract 
\begin{equation}
 f^{exp}_{D_+} =  205.8(8.5)(2.5) \, {\rm MeV}   \,,
\end{equation}
which is in rather good agreement with the HPQCD values in
(\ref{eq:lat}). It should be added that the prediction of
$f_{D_+}$ is in principle more challenging than $f_{D_s}$ 
as the former contains a (chiral) $u$ quark which is harder to
simulate on the lattice than the $s$ quark in $f_{D_s}$.
Decay constants can also be determined by QCD sum rules
to a precision of about $10$-$15\%$ which clearly cannot  
compete with the uncertainties of the prediction given
in Eq.~(\ref{eq:lattice}). The results are consistent within 
uncertainties \cite{stone}. Moreover QCD sum rules predictions
have generally been lower than lattice predictions for the decay constants $f_{D,D_s,B,B_s}$ and this qualitative pattern
remains for the prediction in Eq.~(\ref{eq:lattice})
and therefore does not constitute a source of doubt on the results.
It is not natural for models of NP to give
rise to enhancement in the second generation $(s,c)$
but none in the first family  $(u,d)$.
In reference \cite{Andreas} non-standard
interactions in $S,T,U$-channel were proposed:
$W'$/charged Higgs,
leptoquarks with charges $+2/3$ and $-1/3$.  In principle the phenomenon could also be explained 
by an enhancement of $|V_{\rm cs}| \sim 1.1$, which would contradict 
unitarity in many models. Let us finish this section by noting that 
in principle the recently proposed Lee-Wick Standard Model allows for CKM elements larger than one \cite{KUZ} without violating unitarity. In practice though, the effect can only be significant in the top elements because of the possible closeness of the top mass to a NP scale.

\subsection*{New phase in $B_s$ oscillations?}
Neutral meson oscillations are a fascinating phenomenon
often setting strongest constraints
on NP models in the flavour sector and 
are a realization of EPR correlations. 
Of the four possible oscillation modes $K_0, B_d, B_s, D_0$ 
the third and fourth were only established in the last two years.
If a final state $f$ can be reached from both meson and anti meson,
e.g. $B_s \to f \leftarrow \bar B_s$,
the mixing can be observed in time dependent decay rate. 
In order to give the reader an idea, we shall display a rate
$$
\Gamma(B_s(\bar B_s) \to f) 
\sim \cosh({\frac{\Delta \Gamma_s t}{2}) -  \cos(\phi_s) \sinh(\frac{\Delta \Gamma_s t}{2}})  \pm 
 \sin(\phi_s) \sin(\Delta M_s t)  \, ,
$$
for a case with no relative strong phases and
where the ratio of mixing coefficient $p$ and $q $ is $|p/q|\simeq1$ \cite{PDG}. 
The latter is the case in the SM and in nature \cite{PDG}.
Needless to say that the sign difference in the $\sin(\phi_s)$ term is crucial 
in connection with $B_s$-tagging.
The mass and width difference $\Delta M_s, \Delta \Gamma_s$
and the relative phase difference 
between the mass and width transition elements $\phi_s \equiv - \arg(\Gamma_{12}/M_{12})$  are observables. The phase $\phi_s$, which is often 
called mixing phase,  is an unambiguous signal of 
CP-violation and is predicted to be 
\begin{equation}
\phi_s^{SM} \simeq - 2 \beta_s \equiv   2 \arg[ V_{\rm ts}V_{\rm tb}^* / (V_{\rm cs} V_{\rm cb}^*)] \simeq - 2 \lambda^2 \eta \simeq - 0.04(1) \simeq - 2.0(5)^\circ \,,
\end{equation}
in the SM. It is an example of a good observable as the 
prediction is ``very clean" and it is highly 
sensitive to NP. 
Measurements
of $\Delta M_s$ and $\Delta \Gamma_s$ do not indicate deviation from the SM at the current experimental or theoretical precision
\cite{HFAG}. 

Recently D0 \cite{newD0} and CDF \cite{newCDF} have
published results from time-dependent angular analysis on the 
tagged decay $B_s \to J/\Psi \phi$,
which allows constraints to be set 
on $\Delta \Gamma_s$ and $\phi_s$. 
These results were combined by the UTfit-collaboration \cite{utfit}
with previous measurements. The result is parameterized in terms
of the NP phase $\phi_{B_s}$ 
($\phi_s  \equiv \phi_s^{SM} + 
2 \phi_{B_s}$ \footnote{We  use the sign convention for $\phi_s$ 
from CDF and D0, which is opposite to the one by the UTfit collaboration.}) with the fit result in the Bayesian approach of 
\begin{equation}
{\rm UTfit }\, \mbox{\cite{utfit}} : \quad  \phi_{B_s} = - 19.9(5.6)^\circ  \, .
\end{equation}
Out of a two-fold ambiguity we have quoted the result
which is consistent with information on 
strong phases from $B \to J/\Psi K^*$ by invoking the approximate SU(3)$_F$ symmetry \footnote{
The SU(3)$_F$ limit might not work too well 
because the $\phi$ contains a singlet component contrary to the $K^*$.}. The first result deviates 
by $3.7\sigma$ from the SM\,\footnote{The first paper pointing towards a
discrepancy in the global $B_s$ system within a global analysis 
is reference\,\cite{LenzNierste}. The authors
found a  $2\sigma$ 
discrepancy (without the tagged analysis).}
At a recent workshop the UTfit collaboration has given an 
update of their results, using new partial results of 
D0 dropping an SU(3)$_F$ assumption, resulting in 
$\phi_{B_s} = -18(7)^\circ$ and a $2.7\sigma$ deviation from the SM \cite{capri}.
At the very same workshop CKM-fitter has presented a preliminary analysis, assuming Gaussian distributions 
in the absence of more precise information of 
D0\;\footnote{According to them this tends to underestimate the errors.}, with
a $2.7\sigma$ effect of NP in $\phi_s$
\cite{capri}.
It should be added that the combination of the two results is 
not straightforward since D0 uses information 
on strong phases from $B \to K^* J/\Psi$ and 
SU(3)$_F$ and CDF does not.
In particular D0 and CDF have 
not yet published the combination of their results
but are of course working on it.

\subsection*{``Large'' difference of CP asymmetries 
in $B \to K \pi$ in charged and neutral mode?}
The non-leptonic decay $B \to K \pi$ is a loop 
dominated decay and therefore sensitive to NP.
There has been a tension at
the $2\sigma$ level between ratios of branching fraction 
in charged and neutral modes.
The focus has moved to the CP-asymmetries in that mode \cite{baeklondon}
\begin{eqnarray}
\label{eq:asym}
A_{\rm CP}(B^0 \to K^+ \pi^-)   &=& -0.097(12) \,, \, 
A_{\rm CP}(B^+ \to K^+ \pi^0)   = 0.050(25) \quad \nonumber 
\\[0.2cm]  &\Rightarrow&  
\Delta A_{\rm CP} \neq 0 \;\; @\, 5.3\sigma \, ,
\end{eqnarray}
as emphasized in two recent articles in Nature by BaBar and
Belle \cite{Nature}.
It is worth mentioning that the results are averages of measurements
by BaBar, Belle, CDF \& CLEO in the neutral mode and BaBar \& Belle in the charged mode and that the individual measurements
are all consistent with each other.
Again in order to give the reader an idea we write down the topological
decomposition,
\begin{eqnarray}
{\cal A}(B^0 \to K^+ \pi^-) &=& -T e^{i\gamma} - P  \nonumber \\
\sqrt{2} \, {\cal A}(B^+ \to K^+ \pi^0) &=& 
-(T +C+  A) e^{i\gamma} - P - 
P_{EW} \,,
\end{eqnarray}
in the isospin limit which relates certain amplitudes
with each other.
The symbols $T$ and $C$ denote Cabibbo suppressed tree graphs and 
$C$ stands for color suppressed, $P$ is the QCD emission 
\& annihilation penguin, $P_{EW}$ is the electroweak penguin
and $A$ is the tree annihilation graph. 
In this decomposition we have omitted the color suppressed electroweak penguin
since it is not crucial for the essence 
of the argument. A large difference in the CP 
asymmetries could be caused by 
$C,A$ or $P_{\rm EW}$.
The issue is that the contributions of $C$ and $A$ are
difficult to estimate precisely within QCD and $P_{\rm EW}$
is sensitive to NP, since the amplitude for a hypothetical quark
grows with $m_t'^2/m_Z^2$ due to the famous GIM enhancement effect.

The crude estimate of the hierarchies from naive
factorization $|P| : |T|,|P_{\rm EW}| : |C| \simeq 1 : \lambda : \lambda^2$ with $\lambda \simeq 0.2$  
was given a long time ago \cite{hierarchy}.
To estimate $|A|$ is more difficult but it can for instance 
be argued that the annihilation of the $B$-meson 
is proportional to its wave function which is known
to be small from quark models. In QCD factorization (leading heavy quark limes) the annihilation term is  color suppressed
and only the small 'tree' Wilson coefficient contributes;
From the tables in \cite{BBNS} we infer $|A/T| \lesssim 0.1$ 
\footnote{This ratio was estimated to be $O(\lambda^2)$ 
prior to QCD factorization, e.g. \cite{hierarchy}.}.
Altogether this suggests that the two CP asymmetries in 
\eqref{eq:asym}
are small ($O(\lambda)$ 
times a strong phase suppression) and very close to each other.
The former is the case but not the latter and this is at the heart of the ``new $B \to K \pi$ puzzle''.

A closer look appears appropriate.
Fits of general isospin parameterization 
in $B \to K \pi$ to all observables (branching ratios, 
(time-dependent) CP asymmetries) were undertaken 
recently in the neutral mode in  \cite{cern} and in the charged and
neutral modes in  \cite{FJM}. 
In order to reduce the number of parameters below the number 
of observables \footnote{In the total 
$B \to K \pi$ (isospin limit) system   
there are 11 parameters exceeding the 9 observables.},
quantitative results including error estimates
have been taken into account, e.g. ratio of $P_{\rm EW}/(C+T)$ from QCD factorization\,\cite{BBNS}. 
In reference\,\cite{FJM} the fits constrain 
$|(C+A)/(T-A)|$  to the
range of $[0.52,3.0]$ with a central value $0.89$. Whereas this certainly challenges the old estimates  $|C/T| \sim 0.2, |A/C| \sim 0.2$\,\cite{hierarchy}, the situation in QCD factorization appears less pressing;
$|C/T| \sim 0.28^{+0.3}_{-0.2}$ at NLO\;\cite{sebastian}  and $|A/T| \lesssim 0.1$\,\cite{BBNS}  ($C$ and $A$ add mostly constructive in QCD factorization). The first result comes
with large, and presumably conservative, parametric 
uncertainty  and the latter is difficult to judge because of its model dependence due to endpoint divergences. 
In summary these results
allow for speculations but not
for conclusions. In reference 
\cite{sebastian} a different point was emphasized.
Even in the case where the strong phases are tuned
to satisfy Eq.~\eqref{eq:asym},
the time dependent CP asymmetry $S_{\pi_0 K_s}$
still differs from the SM by $2\sigma$. 
The same qualitative statement can also be inferred
from the explicit formulae in section 3.5.1. of reference \cite{FJM}.

If it shall turn out that there is really a discrepancy then
an enhanced electroweak penguin 
$P_{\rm EW}$ would constitute a primary suspect for NP
contributions.
In this connection the fourth generation was emphasized 
\cite{hou}, because of the $m_t'^2/m_Z^2$.
It was argued that enhanced $P_{\rm EW}$
could accommodate the large CP asymmetry as well as
the deviation in the measurement of the $B_s$ mixing phase
and be of primary interest for electroweak Baryogenesis.

\subsection*{Conclusions}

Of the three puzzles  presented here, the $\phi_s$ mixing phase will presumably be resolved earliest as the LHCb experiment will determine the phase to a precision of $\sim 2^\circ$
for a nominal luminosity in one year. In connection with
the decay constant $f_{D_s}$ it would certainly be desirable to have 
a verification of the result by groups using other methods 
of simulating QCD on the lattice. On the experimental side the  
BES-III experiment will determine $f_{D_s}$ with an 
uncertainty of about $2\%$ \cite{besIII}, 
meeting the theoretical precision advocated by HPQCD \cite{hpqcd},
after four years of operation at nominal luminosity 
of $5\,{\rm fb}^{-1}/{\rm year}$.
Experimental progress on $\Delta A_{\rm CP}(K\pi)$ puzzle,
mostly needed in $K^+\pi^0$ channel (\ref{eq:asym}), will
mainly come from $e^+e^-$ machines since the hadronic machines
as the LHCb are unsuitable to detect a $\pi^0$ in terms of two photons.
As already stated before, deviations in $\phi_s$ and $A_{B \to \pi K}$
would contradict the popular MFV hypothesis and therefore be 
of utmost importance.

We would like to end this write-up by mentioning that 
in choosing three examples we have made choices and
have for instance left out the tension 
between the $\sin(\beta_{\rm eff})$ determination from 
the tree dominated decay $B \to J/\Psi K_s ( b \to c \bar c s)$ and 
penguin dominated decay $B \to \phi K_s  ( b \to s  \bar s s)$ \cite{HFAG}. Recently  constraints
on $\sin(2 \beta)$ were obtained from $\Delta M_s/\Delta M_s$ and $\epsilon_K$ (improved lattice result) and with and without 
$|V_{\rm ub}/V_{\rm cb}|$, which deviate in 
$\sin(2\beta)$ from the two decay types above in
by $1.8$-$2.7\sigma$ \cite{Soni}.
The status of $(g$-$2)_\mu$ has been discussed at a dedicated session at the conference\,\cite{g-2}.

\subsection*{Acknowledgments}
I would like to thank the organizers and
participants of Moriond QCD 2008 for a pleasant
conference.  I am grateful to Thorsten Feldmann, Sebastian J\"ager, Martin Jung, Christopher Smith, Michael Spranger, Sheldon Stone  
and many other colleagues for discussions or comments.
Apologies for any  omitted references.

\end{document}